\documentclass[pre,onecolumn,11pt,aps,amsmath,amsfonts,floatfix,a4paper]{revtex4-1}

\usepackage{txfonts}

\usepackage{amsmath}
\usepackage{hyperref}
\usepackage{graphicx}
\usepackage{physics} 
\usepackage{siunitx}

\newcommand{\ee}{\mathrm{e}}
\newcommand{\ii}{\mathrm{i}}

\newcommand{\scZ}{\mathcal{Z}}
\newcommand{\scT}{\mathcal{T}}

\newcommand{\goH}{\mathfrak{H}}
\newcommand{\scH}{\mathcal{H}}
\newcommand{\goC}{\mathfrak{C}}
\newcommand{\goR}{\mathfrak{R}}
\newcommand{\goBv}{\boldsymbol{\mathfrak{B}}}

\newcommand{\scP}{\mathcal{P}}
\newcommand{\scC}{\mathcal{C}}

\newcommand{\rv}{\boldsymbol{r}}
\newcommand{\Sv}{\boldsymbol{S}}
\newcommand{\hv}{\boldsymbol{h}}

\newcommand{\uv}{\boldsymbol{u}}
\newcommand{\Mv}{\boldsymbol{M}}
\newcommand{\qv}{\boldsymbol{q}}
\newcommand{\zerov}{\boldsymbol{0}}

\newcommand{\Mz}{M_z}
\newcommand{\uvz}{\uv_z}

\newcommand{\nhv}{\hat{\boldsymbol{n}}}

\newcommand{\rmU}{\mathrm{U}}

\newcommand{\rnn}{r\sub{nn}}
\newcommand{\DTO}{\(\mathrm{Dy}_2\mathrm{Ti}_2\mathrm{O}_7\)}
\newcommand{\HTO}{\(\mathrm{Ho}_2\mathrm{Ti}_2\mathrm{O}_7\)}

\newcommand{\beq}[1]{\begin{equation}\label{#1}}
\newcommand{\eeq}{\end{equation}}
\newcommand{\refeq}[1]{Eq.~(\ref{#1})}

\newcommand{\refcite}[1]{Ref.~\cite{#1}}
\newcommand{\refcites}[1]{Refs.~\cite{#1}}
\newcommand{\reffig}[1]{Fig.~\ref{#1}}
\newcommand{\refsec}[1]{Section~\ref{#1}}

\newcommand{\punc}[1]{\,{\text{#1}}}
\newcommand{\sub}[1]{_{\text{#1}}}

\newcommand{\super}[1]{^{\text{#1}}}

\usepackage{color}

\begin{document}

\title{Mean-field theory for confinement transitions and magnetization plateaux in spin ice}
\author{Stephen Powell}
\affiliation{School of Physics and Astronomy, The University of Nottingham, Nottingham, NG7 2RD, United Kingdom}

\begin{abstract}
We study phase transitions in classical spin ice at nonzero magnetization, by introducing a mean-field theory designed to capture the interplay between confinement and topological constraints. The method is applied to a model of spin ice in an applied magnetic field along the \([100]\) crystallographic direction and yields a phase diagram containing the Coulomb phase as well as a set of magnetization plateaux. We argue that the lobe structure of the phase diagram, strongly reminiscent of the Bose--Hubbard model, is generic to Coulomb spin liquids.
\end{abstract}

\maketitle

\section{Introduction}

Classical spin liquids \cite{Balents2010}, such as the Coulomb phase \cite{Henley2010} in spin ice \cite{Bramwell2001} and related systems, are examples of phases whose behavior is not captured by the standard Landau picture of broken symmetries \cite{Landau1999}. Their two defining characteristics are fractionalization, the emergence of excitations not constructed from finite combinations of the elementary degrees of freedom, and topological order, the presence of structure that can only be discerned by observing the system globally \cite{Castelnovo2012}.

In the case of the classical spin ices \cite{Bramwell2001}, a family of magnetic pyrochlore oxides, the consequences of these properties have been of sustained interest, from both theoretical and experimental perspectives. The spins in these materials carry large magnetic moments, and their fractional excitations take the form of magnetic monopoles \cite{Castelnovo2008}, acting as sources for the physical magnetic field. Furthermore, transitions between phases where these excitations are confined and deconfined have particularly interesting properties that cannot be captured by the Landau--Ginzburg theory of phase transitions \cite{MonopoleScalingPRL,MonopoleScalingPRB}.

Spin ice also has the unusual feature that the magnetization is a topologically constrained quantity, and so topological order can be probed directly \cite{Castelnovo2012}. In particular, within the low-energy configuration space relevant at low temperatures, any local dynamics conserves the uniform magnetization \cite{Henley2010}. This fact is known to have interesting consequences for critical properties at certain confinement transitions, such as the Kasteleyn transition in an applied magnetic field \cite{Jaubert2008,SpinIceCQ}.

In this work, we investigate the interplay between these two aspects of the Coulomb spin liquid by studying confinement phase transitions in spin ice across the full range of magnetization. As noted by Henley \cite{Henley2010}, one can distinguish three categories of phase transition from the Coulomb phase: The first, where the magnetization is zero across the transiton, was studied systematically in \refcite{SpinIceHiggs}. A second, where the ordered state has saturated polarization, was considered in \refcites{Jaubert2008,SpinIceCQ}. Here we consider the phase structure and transitions more generally, including the third case, where the magnetization is nonzero but unsaturated, and may or may not change across the transition.

To do so, we introduce a mean-field theory (MFT) designed to capture the distinction between confined and deconfined phases. Standard mean-field approaches, based on the Landau picture of ordered phases, are not well suited to describing confinement transitions. Instead, we take inspiration from the well-known MFT for the Bose--Hubbard model \cite{Fisher1989}, making use of a mapping from spin ice to a quantum model of bosons \cite{SpinIceCQ}. We apply the method to a simplified model of classical spin ice, in order to illustrate the general approach and its physical interpretation. We also briefly discuss extensions to the method that could be used to treat more physically realistic perturbations.

In addition, we present arguments for the generality of our results, beyond mean-field theory and our particular choice of Hamiltonian. We argue that the phase diagram generically consists of a set of confined phases at low temperature in which the magnetization is, to a very good approximation, fixed. This plateau structure is strongly reminiscent of the lobes present in the phase diagram of the Bose--Hubbard model \cite{Fisher1989,Sachdev2011}, a connection that helps clarify the general phase structure.

While the focus here is on classical spin ice in a \([100]\) field, the distinction between transitions in different flux sectors is more general. For example, spin ice in a field along the \([111]\) direction also exhibits magnetization plateaux \cite{Matsuhira2002,Moessner2003,Isakov2004}, while it has been suggested that the zero-field ground state of quantum spin ice may have nonzero magnetization \cite{Shannon2012}.

In \refsec{SecModel}, we introduce the model to be used and briefly outline the general properties of the Coulomb phase and confinement transitions. The mean-field method is then described and applied to the model in \refsec{SecMeanFieldTheory}. We present general arguments for the phase structure and the nature of the phase transitions in \refsec{SecPhaseStructure}, before concluding in \refsec{SecConclusions}.

\section{Model and basic properties}
\label{SecModel}

Our presentation will be based on the case of classical spin ice in a magnetic field applied along the \([100]\) crystallographic direction. We first introduce the nearest-neighbor model of spin ice (NNSI) that describes the physics of the Coulomb spin liquid, before discussing perturbations, such as an applied field, and the resulting ordering transitions.

\subsection{Nearest-neighbor spin ice}
\label{SecNNSI}

\begin{figure}
\begin{center}
\includegraphics[width=0.35\textwidth]{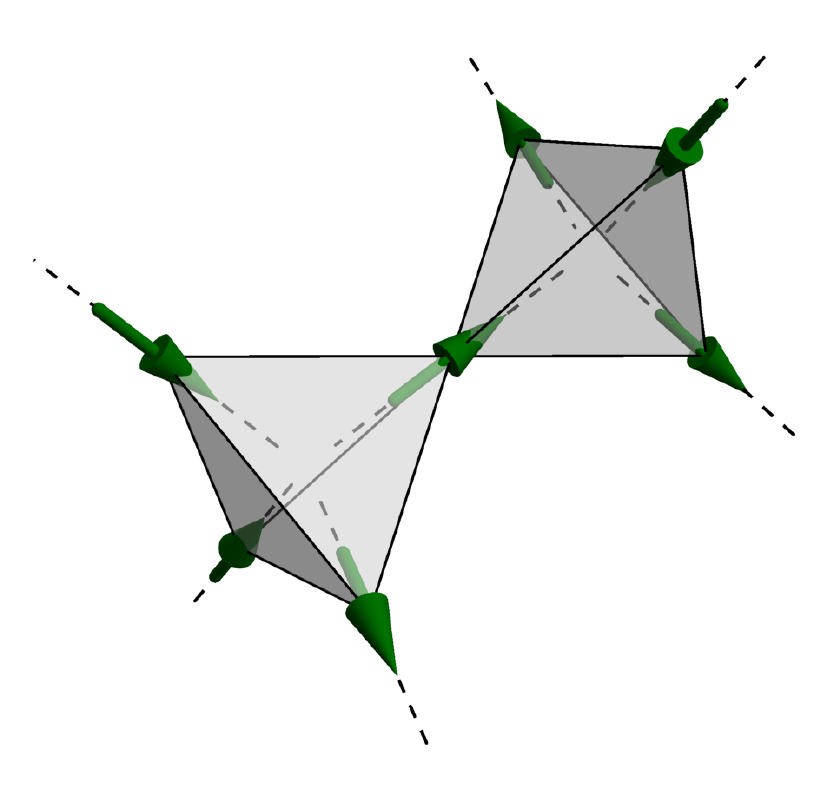}
\end{center}
\caption{A pair of tetrahedra forming part of the pyrochlore lattice. In the classical spin-ice compounds, magnetic moments \(\Sv_i\) (shown as green arrows) lie at the sites \(i\) of the lattice and are constrained to point along the local \(\langle 111 \rangle\) axes (dashed lines) that join the centers of adjacent tetrahedra.}
\label{FigTetrPair}
\end{figure}

The spin ices \cite{Bramwell2001} are a family of frustrated magnetic materials with moments on the sites \(i\) of a pyrochlore lattice, a network of corner-sharing tetrahedra. Prominent examples are the ``classical'' spin ices, such as \DTO\ and \HTO, which are well described in terms of classical spins \(\Sv_i\). Each spin is subject to strong easy-axis anisotropy along the local \(\langle 111 \rangle\) direction \(\nhv_i\) which joins the centres of the two tetrahedra sharing site \(i\), as shown in \reffig{FigTetrPair}.
The anisotropy results from a large (of order \SI{100}{K}) crystal-field splitting between states with maximum projection along \(\nhv_i\) and all other states in the single-moment Hilbert space. This large gap ensures that tunneling processes between the low-energy states are strongly suppressed, rendering the moments effectively classical \cite{Bramwell2001}. For the low temperatures of interest, it is therefore sufficient to treat the spins as effectively constrained to \(\Sv_i = \sigma_i \nhv_i\), where \(\sigma = \pm 1\) is a classical Ising variable.

Each tetrahedron \(t\) in the pyrochlore lattice can be labeled according to its orientation \(\epsilon_t = \pm 1\), and the lattice structure is such that all neighbors of \(t\) have orientation \(-\epsilon_t\). The fixed unit vector \(\nhv_i\) for each pyrochlore site \(i\), directed along the line joining the centers of its two tetrahedra, can be chosen to point towards the tetrahedron with \(\epsilon_t = -1\), and so \(\epsilon_t\sigma_i\) indicates whether \(\Sv_i\) points out of (\(+1\)) or into (\(-1\)) tetrahedron \(t\). With this convention, \(\nhv_i\cdot\nhv_j = -\frac{1}{3}\) and hence \(\Sv_i\cdot\Sv_j = -\frac{1}{3}\sigma_i\sigma_j\), for all nearest neighbors \(i\) and \(j\).

The minimal Hamiltonian that captures the spin-liquid physics of the classical spin ices \cite{Castelnovo2012} contains only interactions between nearest-neighbor pairs \(\langle i j \rangle\),
\begin{equation}
\scH\sub{nn} = -J \sum_{\langle i j \rangle} \Sv_i \cdot \Sv_j = \frac{J}{3} \sum_{\langle i j \rangle} \sigma_i \sigma_j\punc{.}
\end{equation}
The coupling between moments \(\Sv_i\) is ferromagnetic (\(J > 0\)), incorporating the net effect of the dipolar interactions between nearest neighbors and (antiferromagnetic) exchange \cite{Bramwell2001}. The effective interactions between the Ising variables \(\sigma_i\) are therefore antiferromagnetic, and hence frustrated. Further-neighbor interactions, both dipolar and exchange, are significant in spin ice, but can be treated as a relatively small perturbation to \(\scH\sub{nn}\) \cite{Castelnovo2012}; we will return to these in \refsec{SecPerturbations}.

Because two sites are nearest neighbors if and only if they share a tetrahedron, the interactions can be rewritten as
\begin{equation}
\scH\sub{nn} = \frac{2}{3}J \sum_t \left(Q_t^2 - 1\right)\punc{,}
\end{equation}
where \(\sum_t\) includes all tetrahedra (of both orientations) and
\begin{equation}
Q_t = \frac{1}{2}\epsilon_t \sum_{i \in t} \sigma_i
\end{equation}
is referred to as the (fictional) ``charge'' on tetrahedron \(t\) in terms of a sum over its sites \(i \in t\). The energy is therefore minimized by configurations with \(Q_t = 0\), which are those where two spins point into and two out of each tetrahedron. Configurations that satisfy this condition are said to obey the {\it ice rule} \cite{Bramwell2001,Castelnovo2012}; all configurations that do so at every tetrahedron are degenerate minimal-energy states of \(\scH\sub{nn}\).

Tetrahedron configurations with three spins in and one out, or vice versa, have energy \(\Delta = +\frac{2}{3}J\) above the ice-rule configurations; they have charge \(Q_t = \pm 1\) and are hence referred to as ``monopoles''. (Note that \(\sum_t Q_t = 0\), and so all configurations are globally charge-neutral.) For \DTO, the effective nearest-neighbor coupling is \(\frac{1}{3}J = \SI{1.1}{K}\) \cite{Bramwell2001}, corresponding to a monopole cost of \(\Delta = \SI{2.2}{K}\). The corresponding Boltzmann weight \(\zeta = \ee^{-\Delta/T}\) is called the monopole fugacity. While the significance of the term ``monopole'' is here only that they carry effective ``charge'' \(Q_t\), it has been argued that monopole excitations in the classical spin ice materials in fact carry true magnetic charge \cite{Castelnovo2008}.

For temperature \(T\) significantly below \(\Delta\), the system is effectively constrained by the ice rule, and the ensemble of ice-rule configurations therefore determines its behavior \cite{Bramwell2001}. The number of such configurations grows exponentially with the total number of spins \(N\), and so the entropy density remains nonzero even well below \(\Delta\), the temperature scale at which an unfrustrated system would order. In spite of the absence of order, the ensemble exhibits correlations that decay only algebraically with distance. The resulting correlated paramagnet is referred to as a Coulomb phase \cite{Henley2010}, and is an example of a classical spin liquid \cite{Balents2010}.

The long-wavelength properties of the Coulomb phase can be captured by coarse-graining the vector spins \(\Sv\) to give a continuum field \(\goBv(\rv)\) referred to as the effective ``magnetic field'' \cite{Henley2010}. The ice-rule constraint applied to \(\Sv\) implies that \(\goBv\) is divergenceless, while monopoles act as sources or sinks, according to an effective Gauss law. The resulting predictions for spin--spin correlation functions are in good agreement with neutron-scattering experiments in classical spin ice \cite{Henley2010}.

The ice-rule configurations also display interesting topological properties \cite{Castelnovo2012}. For our purposes, the most important is that any pair of ice-rule configurations is connected by flipping all spins along one or more loops. For a system with periodic boundary conditions (which we will assume), such a loop can only change the total magnetization \(\Mv = \sum_i \Sv_i\) if it has nontrivial winding number. For other systems that host Coulomb phases \cite{Henley2010}, such as classical dimer models, an analogous quantity, called the ``flux'', can be defined, but does not necessarily correspond to a quantity that is so directly accessible experimentally.

\subsection{Confinement transitions}
\label{SecConfinementTransitions}

A second important property of the Coulomb phase is that monopoles, defects in the ice rule, are deconfined \cite{Castelnovo2008}. To make this statement precise, imagine imposing a pair of monopoles with opposite charges \(\pm 1\) at tetrahedra \(t\) and \(t'\) in a background that otherwise obeys the ice rule. Let \(\goC_{t,t'}\) denote the set of spin configurations \(\{\sigma_i\}\) compatible with this charge configuration. Their number \(\scZ\sub{m}(\rv_{t,t'}) = \lvert\goC_{t,t'}\rvert\) is a function of the separation \(\rv_{t,t'}\) that decreases with \(\lvert \rv_{t,t'}\rvert\) but reaches a finite limit as \(\lvert \rv_{t,t'}\rvert \rightarrow \infty\). The result is an effective entropic interaction
\begin{equation}
U\sub{m}(\rv_{t,t'}) = -T\ln \scZ\sub{m}(\rv_{t,t'})
\end{equation}
between the charges that is bounded, and hence allows them to be separated to infinity at finite free-energy cost. In fact, since monopoles are charges in the continuum field \(\goBv\), coarse-graining predicts that the interaction obeys the Coulomb law, \(U\sub{m}(\rv) \sim \text{const} - \lvert \rv \rvert^{-1}\), for large \(\lvert\rv\rvert\) \cite{Henley2010}.

More generally, consider the partition function
\begin{equation}
\scZ\sub{m}(\rv_{t,t'}) = \sum_{\{\sigma_i\} \in \goC_{t,t'}} \ee^{-V/T}\punc{,}
\end{equation}
where \(V\) is a perturbation that splits the degeneracy of the ice-rule configurations. This will tend to suppress fluctuations, and can eventually, as the temperature is reduced compared to the scale of the perturbation, drive the system out of the Coulomb phase. A transition into a phase where \(U\sub{m}(\rv_{t,t'})\) increases without limit as \(\lvert \rv_{t,t'}\rvert \rightarrow \infty\), and so \(\scZ\sub{m}(\rv_{t,t'})\rightarrow 0\), is referred to as a confinement transition \cite{MonopoleScalingPRL,MonopoleScalingPRB}. (In terms of the effective continuum field \(\goBv\), Gauss' law implies that an imposed pair of charges must be joined by a fixed quantity of flux. Linear confinement, where \(U\sub{m}(\rv) \sim \lvert\rv\rvert\) for large \(\rvert\rv\rvert\), occurs when the flux forms a narrow tube connecting the monopoles, with finite tension.)

While confinement may occur simultaneously with appearance of conventional symmetry-breaking order \cite{Sreejith2014}, there is no local order parameter that provides a signature of confinement. The confinement criterion, that \(\scZ\sub{m}(\rv_{t,t'})\) approaches zero as \(\lvert \rv_{t,t'}\rvert \rightarrow \infty\) in a confined phase and conversely has a finite limit in a deconfined phase, instead resembles the characterization of a conventional symmetry-broken phase in terms of long-range order. This connection can be pursued further by considering the partition function with an unconstrained charge distribution and a position-dependent fugacity \(\zeta_t\),
\begin{equation}
\scZ[\zeta_t] = \sum_{\{\sigma_i\}} \ee^{-V/T} \prod_t \zeta_t^{\lvert Q_t \rvert}\punc{,}
\end{equation}
in terms of which \cite{FootnoteTotalChargeZero}
\begin{equation}
G\sub{m}(\rv_{t,t'}) \equiv \frac{\scZ\sub{m}(\rv_{t,t'})}{\scZ} = \frac{1}{2}\left. \frac{\partial}{\partial \zeta_t} \frac{\partial}{\partial \zeta_{t'}} \ln \scZ[\zeta_t] \right\rvert_{\zeta_\cdot = 0}\punc{,}
\label{EqMonopoleDistributionFunction}
\end{equation}
where \(\scZ = \left.\scZ[\zeta_t]\right\rvert_{\zeta_\cdot = 0}\) is the partition function restricted to ice-rule configurations. The relationship between the monopole distribution function \(G\sub{m}\) and the fugacity \(\zeta\) expressed by \refeq{EqMonopoleDistributionFunction} is exactly analogous to the fluctuation--dissipation theorem relating, for example, a spin--spin correlation function and an applied field. This connection motivates a mean-field theory, analogous to standard Weiss theory \cite{Yeomans1992}, that describes the confinement transition in terms of an effective self-consistent fugacity.

It should be noted that the confinement distinction is only sharp in the limit \(\zeta = 0\), because any nonzero density of monopoles screens the effective interaction \(U_{+1,-1}\) at large separation. There are nonetheless implications for the critical properties away from this limit \cite{MonopoleScalingPRL,MonopoleScalingPRB}. If the phase transition also involves a different type of order (such as spontaneous symmetry breaking), then it may survive for \(\zeta > 0\), but with a different universality class \cite{Sreejith2014}.

\subsection{Perturbations to NNSI}
\label{SecPerturbations}

Within the nearest-neighbor model \(\scH\sub{nn}\), all configurations that obey the ice rules are degenerate and the system exhibits a deconfining Coulomb phase. To drive a confinement transition, one must add perturbations that split the degeneracy of the ice-rule configurations. Here we consider a Hamiltonian of the form
\begin{equation}
\label{EqHamiltonian}
\scH = \scH\sub{nn} + \scH_u + \scH_{\hv}\punc{,}
\end{equation}
where \(\scH_{\hv}\) is the coupling to an applied magnetic field and \(\scH_u\) contains additional interactions described below.

Coupling to a magnetic field \(\hv\) is described by a Zeeman term
\begin{equation}
\scH_{\hv} = - \hv \cdot \Mv = - \hv \cdot \sum_i \Sv_i
\punc{,}
\end{equation}
where a factor dependent on the size of the magnetic moments has been included in the definition of \(\hv\). The case that we consider here has \(\hv = h \uvz\) where \(\uvz\) is a unit vector along \([100]\), and so \(\hv\cdot \Sv_i = h \uvz \cdot \sigma_i \nhv_i = \pm \frac{1}{\sqrt{3}}h\sigma_i\). The coupling can therefore be written as
\begin{equation}
\label{EqZeemanCoupling}
\scH_{\hv} = -\frac{h}{\sqrt{3}} \sum_i \eta_i \sigma_i\punc{,}
\end{equation}
where \(\eta_i = \sqrt{3}\nhv_i \cdot \uvz = \pm 1\). Using the choice described in \refsec{SecNNSI}, where \(\nhv_i\) points towards the tetrahedron with \(\epsilon_t = -1\), \(\eta_i\) alternates in sign on successive \((100)\) planes of spins. The component of the magnetization along the field can likewise be expressed as
\begin{equation}
\Mz = \sum_i \uvz \cdot \Sv_i = \frac{1}{\sqrt{3}} \sum_i \eta_i \sigma_i\punc{.}
\end{equation}

The perturbation \(\scH_{\hv}\) alone can drive a confinement transition, referred to as a Kasteleyn transition \cite{Jaubert2008}. For \(\Delta \gg h \gg T\), all spins align with the field (to the extent allowed by the easy-axis anisotropy), an arrangement that satisfies the ice rules. Any excitation within the ice-rule states involves flipping a loop of spins, but starting from the fully polarized state, the only available loops span the system in the \([100]\) direction. Such excitations therefore have energy cost proportional to the linear system size \(L_z\) and are suppressed in the thermodynamic limit.

The result is a strictly saturated magnetization \(\Mz = M\sub{sat} \equiv \frac{1}{\sqrt{3}}N\) even for nonzero \(T/h\) (but in the limit \(T/\Delta \rightarrow 0\)). Above a temperature \(T\sub{K}\), however, the energy cost is outweighed by the gain in entropy, also \(\propto L_z\), associated with the multitude of available paths for a spanning loop, and a Kasteleyn transition occurs from the saturated paramagnet to the Coulomb phase. The transition is possible only because the ice rule restricts to loop-like excitations; away from the limit \(T/\Delta = 0\), the transition is replaced by a crossover \cite{Jaubert2008,MonopoleScalingPRB}.

\begin{figure}
\begin{center}
\includegraphics[width=0.85\textwidth]{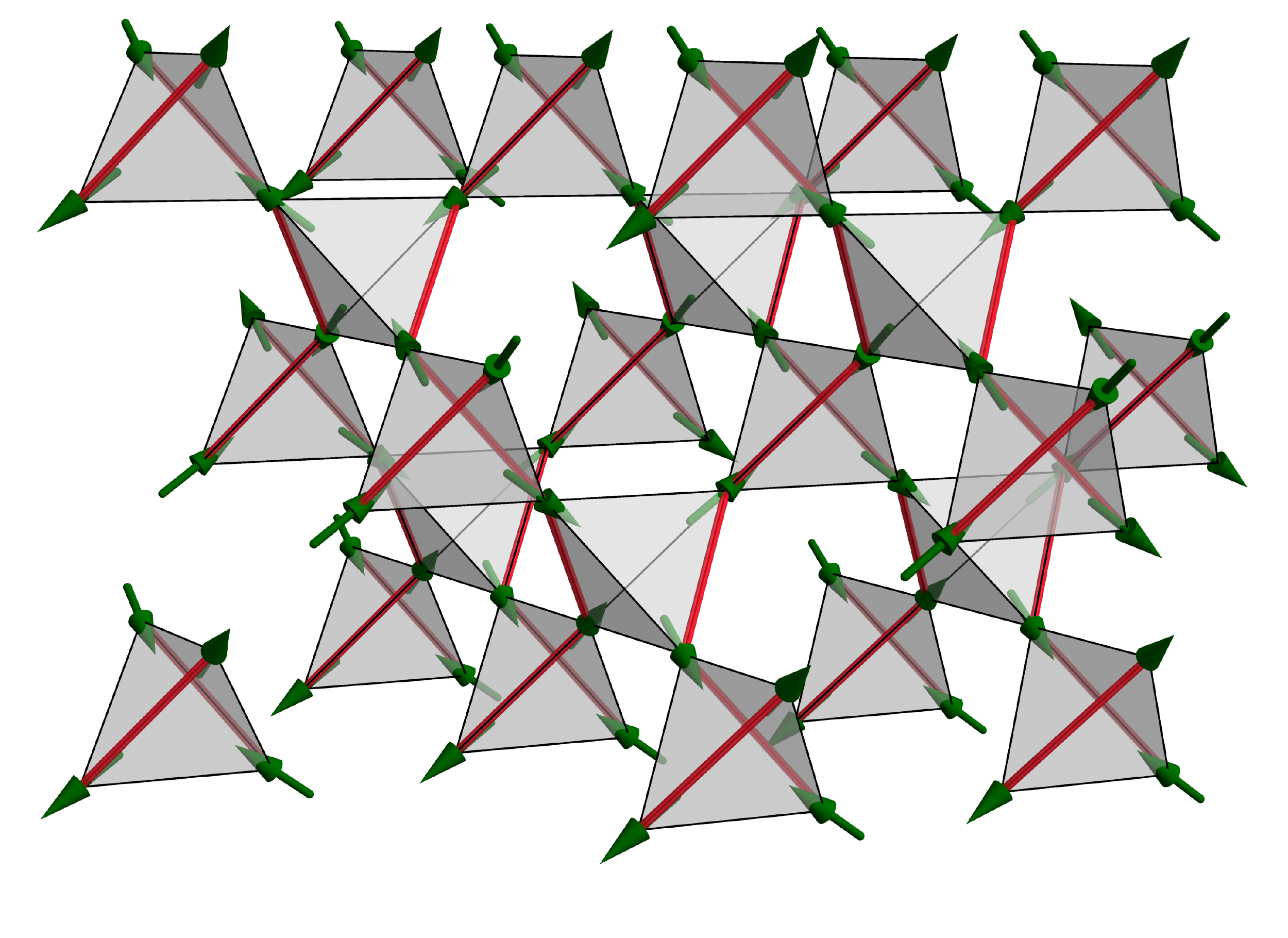}
\end{center}
\caption{Melko--den Hertog--Gingras (MDG) state and a perturbation that favors it. The spins (green arrows) show the fully ordered MDG configuration, which has \(\Mv = \zerov\) and ordering wavevector along \([100]\) (vertical). The spins in each \((100)\) plane are aligned (to the extent allowed by the local easy axes), and between successive planes the magnetization rotates in the right-handed sense by \(\SI{90}{\degree}\). The perturbation \(\scH_u\) reduces the ferromagnetic coupling of the red bonds by \(u\), and hence favors the configuration shown, or the one with all spins inverted.}
\label{FigMDG}
\end{figure}

To study the interplay between the topological constraints on the magnetization, inherent in the ice-rule configurations, and conventional ordering transitions, we also consider an additional interaction \(\scH_u\) that induces spontaneous symmetry breaking. A natural choice would be the further-neighbor coupling present in the spin-ice compounds, due to a combination of further-neighbor exchange \cite{Henelius2016} and long-range dipolar interactions. The latter have been shown using Monte Carlo \cite{Melko2001} to lead to a phase transition into the ordered configuration shown in \reffig{FigMDG}, which we will call the Melko--den Hertog--Gingras (MDG) state \cite{FootnoteNomenclature}.

\begin{figure}
\begin{center}
\includegraphics{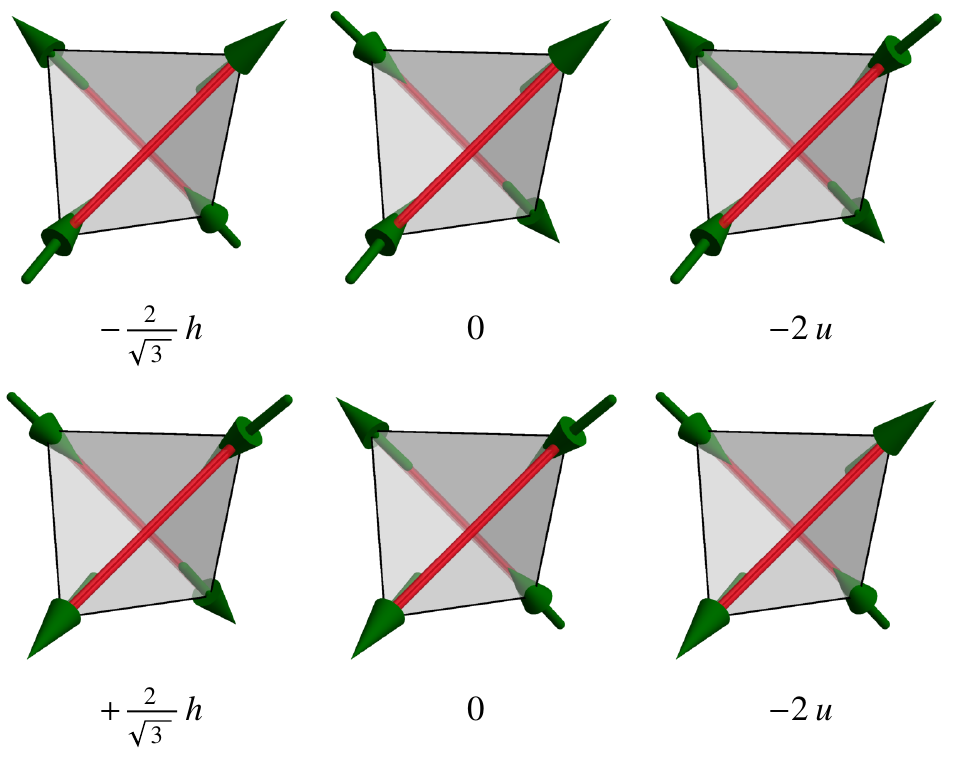}
\end{center}
\caption{The six ice-rule configurations of a tetrahedron and their energies with the perturbation \(\scH_u\), defined in \refeq{EqHu}, as well as an applied field \(h\) along \([100]\) (upwards). The red bonds, as in \reffig{FigMDG}, are those where the ferromagnetic coupling is reduced by \(u\). The configurations favored by this perturbation are therefore those in the right column where the spins joined by the red bond are antialigned.}
\label{FigTetrGrid}
\end{figure}

The actual perturbation \(\scH_u\) that we will use here is one that is simpler to treat within our mean-field theory but that is expected to lead to an ordered phase of the same type. It is given by
\begin{equation}
\scH_u = +\frac{u}{2}\sum_{\langle ij \rangle \in \goR} (3\Sv_i\cdot\Sv_j - 1) = -\frac{u}{2} \sum_{\langle ij \rangle \in \goR} (\sigma_i \sigma_j + 1)\punc{,}
\label{EqHu}
\end{equation}
where \(\goR\) is the set of nearest-neighbor bonds highlighted in \reffig{FigMDG}, which form a set of left-handed screw chains. The effect of \(u > 0\) is to reduce the strength of the interactions on these bonds and to favor the MDG configuration shown, or the one with all spins inverted. \reffig{FigTetrGrid} shows the energy for each of the six configurations of a single tetrahedron that obey the ice rule, including both the applied field and the perturbation \(\scH_u\).

Unlike dipolar interactions, the perturbation \(\scH_u\) reduces the symmetry of the lattice by picking an axis and a chirality. The 12-fold degeneracy of the MDG state is therefore reduced by a factor of \(3 \times 2\), but the symmetry under inversion of all spins remains, and so an ordering transition into this state spontaneously breaks an Ising symmetry. This transition (at \(h = 0\)) is in fact a type considered in \refcite{SpinIceHiggs}, where it was argued to be most likely of first order.

As we will show, this simplified model has a phase diagram that illustrates general features emerging from the interplay between deconfinement, ordering, and topological constraints. In \refsec{SecConclusions}, we will briefly discuss the prospects for including more realistic perturbations in the mean-field theory.

\subsection{Mapping to extended Bose--Hubbard model}
\label{SecBosonMapping}

Some aspects of the phase structure can be elucidated using a mapping between classical spin ice and an effective quantum model of hard-core bosons on a lattice \cite{Jaubert2008,SpinIceCQ}. This proceeds by identifying the \([100]\) direction, along which the external field is applied, with the imaginary-time axis in a path-integral representation of the quantum problem. The thermodynamic limit of the original classical system therefore corresponds to the zero-temperature limit in the quantum problem.

This mapping has previously been studied in detail for the Hamiltonian \(\scH\sub{nn} + \scH_{\hv}\) \cite{SpinIceCQ}. The fully polarized state (along the \(+z\) direction, say) is identified with the vacuum; flipped spins relative to this configuration form system-spanning loops, which are viewed as the world lines of bosons. The result is an effective model of hard-core bosons on a lattice with density \(\rho = \frac{1}{2} ( 1 - \Mz/M\sub{sat} )\) and hence chemical potential \(\mu \sim -h\). In general, one expects all other local terms compatible with symmetry, such as hopping between nearby sites and further-neighbor interactions, with parameters related in nontrivial ways to those of the spin model \cite{SpinIceCQ}.

As \(\mu\) is increased, the system passes from the vacuum through a quantum phase transition \cite{Sachdev2011} to a superfluid, which corresponds to the Coulomb phase of spin ice. (Off-diagonal long-range order in the superfluid can be associated with deconfinement in the Coulomb phase \cite{SpinIceCQ}.) For sufficiently large positive \(\mu\), the system is driven into a fully occupied Mott insulator, with \(\rho = 1\), corresponding to a saturated paramagnet with \(\Mz = -M\sub{sat}\) in an inverted applied field. Critical properties at the phase transitions in spin ice are, {\it mutatis mutandis}, equivalent to those in the quantum model \cite{SpinIceCQ}.

The presence of interactions between bosons on different lattice sites also makes possible Mott insulators with fractional filling. These can be identified with confining phases of spin ice (because they lack superfluid order) in which the magnetization is fixed to a value less than \(M\sub{sat}\). The MDG phase, for example, is an ordered phase with \(\Mz = 0\), and corresponds to a Mott insulator at half filling.

In general, more exotic phases such as supersolids are also possible in the effective quantum model. A supersolid would correspond to a Coulomb phase with broken spatial symmetry; such phases have been proposed in spin ice \cite{BrooksBartlett2014}, but are not captured by the mean-field theory used here.

\section{Transfer-matrix mean-field theory}
\label{SecMeanFieldTheory}

The phase structure of \(\scH\) in \refeq{EqHamiltonian} is expected to contain: the Coulomb phase, characterized by deconfinement of monopoles; the saturated paramagnet, in which fluctuations are suppressed by the ice-rule constraints; and the MDG phase, with conventional order that breaks spin-inversion and spatial symmetries. A mean-field theory that describes all three must therefore capture the confinement--deconfinement distinction, unlike mean-field approaches based on finite clusters \cite{Reimers1991}, and also respect the spatial structure, unlike Bethe-lattice calculations \cite{Jaubert2008}.

Here we apply an approach that is analogous to the standard mean-field theory for the Bose--Hubbard model \cite{Fisher1989}, but is applied directly to the classical partition function, written in terms of a transfer matrix. (For other applications of the variational method to the transfer matrix, see \refcites{Baxter,Rujan}.) We start by partitioning the lattice into chains that span the system, which are the minimal units that can obey the ice rule. An effective model is then defined for each chain, with the coupling between chains treated self-consistently.

In standard mean-field theory for a symmetry-breaking transition \cite{Yeomans1992}, coupling between lattice sites is replaced by an effective field related to the order parameter. Here, in the absence of a local order parameter, we capture the confinement--deconfinement transition by introducing an effective self-consistent fugacity \(\zeta\super{mf}\) for monopoles. Since, according to \refeq{EqMonopoleDistributionFunction}, the monopole distribution function can be interpreted as the correlation function corresponding to a monopole fugacity, a nonzero \(\zeta\super{mf}\) implies deconfinement.

This decoupling is closely analogous to that used for the Bose--Hubbard model, where superfluid order is captured by introducing a self-consistent source term for bosons. Following the mapping between this quantum problem and the classical statistical mechanics of spin ice, we apply the method at the level of the transfer matrix. It should be emphasized that this differs from the standard mean-field approach for classical statistical systems, where one chooses a trial distribution to minimize the free energy. (The latter method cannot be used, because a trial configuration with nonzero fugacity does not enforce the ice rule and therefore has infinite energy.)

While we treat deconfinement using an effective mean-field fugacity \(\zeta\super{mf}\), we set the true monopole fugacity \(\zeta = \ee^{-\Delta/T}\) to zero. Nonzero \(\zeta\) could straightforwardly be included in the approximation and is analogous to standard Weiss mean-field theory in the presence of an applied magnetic field \cite{Yeomans1992}.

\subsection{Decomposition of the pyrochlore lattice into chains}

\begin{figure}
\begin{center}
\includegraphics[width=0.85\textwidth]{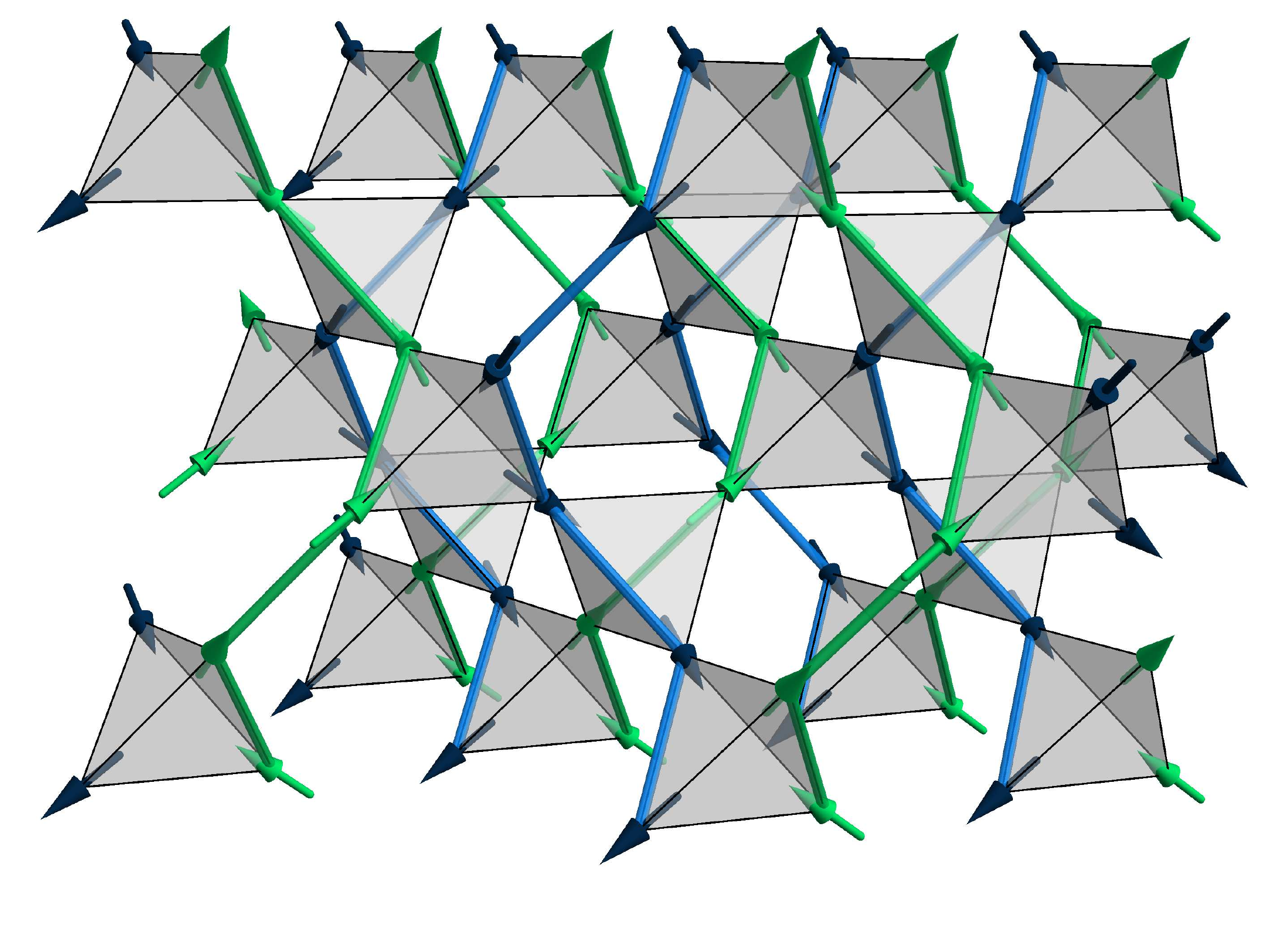}
\end{center}
\caption{Decomposition of the MDG state into screw chains of spins, shown as thick green and blue lines, with axes along the \([100]\) crystallographic direction. In the MDG state, the chains are ordered in a checkerboard pattern, with spins along each chain aligned and neighboring chains antialigned.}
\label{FigMDGChains}
\end{figure}

We choose to decompose the pyrochlore lattice into right-handed screw chains with axes parallel to the \([100]\) direction, as illustrated in \reffig{FigMDGChains}.
For our purposes, this choice has two advantages: First, the chains span the system in the imaginary-time direction under the mapping to a bosonic model, and so the connection to the Bose--Hubbard mean-field theory is transparent. Second, the ordered states that we aim to describe have natural interpretations in terms of chain order: in the MDG state, shown in \reffig{FigMDGChains}, the chains order in a checkerboard pattern, while in the fully polarized paramagnet, all chains are aligned. Other orderings, such as the partially polarized state observed by Lin and Kao \cite{Lin2013}, can be described in terms of ordered configurations of the same chains, while the general approach can likely be adapted for ordered states that are not captured by this decomposition of the lattice. (For example, McClarty et al.\ \cite{McClarty2015} consider ordered states described in terms of linear chains of spins.)

The division of the lattice into screw chains has the property that each chain \(c\) has \(4\) neighbors, which it meets in turn at tetrahedra in successive \((100)\) layers. The set of tetrahedra at a given layer, at vertical position \(z\), amounts to a partition of the chains into pairs, which we denote \(\scP_z\). We also define \(\scC_{c,z}\) as the chain that meets chain \(c\) in a tetrahedron at layer \(z\). The pyrochlore lattice is symmetric under translations by \(\Lambda_z = 4\) layers in the \([100]\) direction \cite{SpinIceCQ}, and so \(\scP_{c+\Lambda_z} = \scP_{c}\) and \(\scC_{c,z+\Lambda_z} = \scC_{c,z}\).

In the quantum mapping, the vertical direction is treated as imaginary time. Applying the mapping at a microscopic level would therefore lead to a time-evolution operator that couples different pairs of lattice sites at successive time steps. This ``stroboscopic'' \cite{Bukov2015} implementation of the hopping in the effective quantum Hamiltonian has been shown to be irrelevant for the critical properties \cite{SpinIceCQ}, but must be handled with some care when implementing the mean-field theory.

\subsection{General approach}

The partition function \(\scZ\) can be expressed exactly in terms of a set of operators acting on degrees of freedom associated with the chains. To each chain \(c\) is associated a pair of basis vectors, \(\ket{\uparrow}_c\) and \(\ket{\downarrow}_c\), representing the orientation, up or down (i.e., \(\sigma_i\eta_i = \pm 1\)), of the spin at a given vertical position \(z\) along the chain. Denoting by \(\goH_c\) the space spanned by these vectors, the full Hilbert space for a layer of the system is \(\goH = \bigotimes_c\goH_c\). (Note that there is a single two-dimensional Hilbert space for each chain, and so the basis vectors are labeled only by their chain \(c\), and not by their position \(z\) along the chain.)

The partition function can then be written as
\begin{equation}
\label{EqPartitionFunction}
\scZ = \Tr \prod_{z=L_z}^1 \scT_z = \Tr \left(\scT_{L_z}\scT_{L_z - 1}\dotsm\scT_{2}\scT_1\right)\punc{,}
\end{equation}
where \(L_z\) is the system size (number of layers) in the \(z\) direction and
\begin{equation}
\label{EqLayerTransferOperator}
\scT_z = \prod_{c c' \in \scP_z} \scT_{cc'}\punc{.}
\end{equation}
is the transfer operator for the layer of tetrahedra at vertical position \(z\). (We assume periodic boundary conditions in the \(z\) direction.) The tetrahedron transfer operator \(\scT_{cc'}\), acting on \(\goH_c\otimes\goH_{c'}\), accounts for the possible configurations of the chains \(c\) and \(c'\) on entering and leaving this tetrahedron. Using the symmetry of the pyrochlore lattice under translations by \(\Lambda_z = 4\) layers in the \(z\) direction, the product in \refeq{EqPartitionFunction} can be rewritten as
\begin{equation}
\scZ = \Tr \left[\scT^{(\Lambda_z)}\right]^{L_z/\Lambda_z}\punc{,}\quad\text{where}\quad\scT^{(\Lambda_z)}=\prod_{z=\Lambda_z}^{1}\scT_z
\punc{.}
\end{equation}
In the thermodynamic limit, the partition function is therefore given by \(\left[\rho\left(\scT^{(\Lambda_z)}\right)\right]^{L_z/\Lambda_z}\), where \(\rho\) is the spectral radius (largest absolute eigenvalue). It should be noted that, even if \(\scT_z\) is hermitian, in general \([\scT_z,\scT_{z'}]\neq 0\) for \(z \neq z' \pmod{\Lambda_z}\), and so \(\scT^{(\Lambda_z)}\) is not hermitian.

The approximation we propose is to replace the system with an effective model of decoupled chains, whose parameters are determined self-consistently. The partition function for chain \(c\) is taken as
\begin{equation}
\scZ\super{mf}_c = \Tr_c \prod_{z=L_z}^{1} \scT\super{mf}_{c,z}\punc{,}
\label{EqPartitionFunctionMF}
\end{equation}
where the effective transfer operator for chain \(c\) is
\begin{equation}
\scT\super{mf}_{c,z} = \bra{v_{\scC_{c,z},z}}_{\scC_{c,z}} \scT_{c,\scC_{c,z}} \ket{v_{\scC_{c,z},z-1}}_{\scC_{c,z}}\punc{,}
\end{equation}
an operator on \(\goH_c\). The normalized vector \(\ket{v_{c,z}}_{c}\in\goH_c\), which is interpreted as an ansatz for the configuration of chain \(c\) immediately above layer \(z\), is determined self-consistently as the thermal state of the one-dimensional system defined by \refeq{EqPartitionFunctionMF}. Assuming a self-consistent solution exists with at least periodicity under translation in \(z\) by \(\Lambda_z\), it is given by the normalized (right) eigenvector with largest absolute eigenvalue of
\begin{equation}
\prod_{z' = z}^{z - \Lambda_z + 1} \scT\super{mf}_{c,z}\punc{.}
\end{equation}

For the types of phases that we aim to describe, it will in fact be sufficient to consider \(\ket{v_{c,z}}_{c} = \ket{v_c}_c\) uniform along each chain. This leads to the simplification that
\begin{equation}
\label{EqTmfc}
\scT\super{mf}_{c,z} = \bra{v_{\scC_{c,z}}}_{\scC_{c,z}} \scT_{c,\scC_{c,z}} \ket{v_{\scC_{c,z}}}_{\scC_{c,z}}\punc{,}
\end{equation}
which is therefore hermitian. For such a solution to exist, \(\ket{v_c}_c\) must be a simultaneous eigenvector of \(\scT_{c,z}\super{mf}\) for all \(z\).

\subsection{Two-chain order in spin ice}

For the case with an applied field \(\hv\) and a chain-favoring interaction \(u\), the tetrahedron transfer operator can be written explicitly as
\begin{equation}
\label{EqTetrahedronTransferOperator}
\scT_{cc'} =
\begin{aligned}[t]
&\ee^{\frac{2}{\sqrt{3}}h/T} \ket{\uparrow}_c \ket{\uparrow}_{c'}\bra{\uparrow}_{c}\bra{\uparrow}_{c'}
+ \ee^{-\frac{2}{\sqrt{3}}h/T}\ket{\downarrow}_c \ket{\downarrow}_{c'}\bra{\downarrow}_c \bra{\downarrow}_{c'}\\
&{}+\ee^{2u/T}\ket{\uparrow}_c\ket{\downarrow}_{c'}\bra{\uparrow}_c \bra{\downarrow}_{c'}
+\ee^{2u/T}\ket{\downarrow}_c\ket{\uparrow}_{c'}\bra{\downarrow}_c \bra{\uparrow}_{c'}\\
&{}+\ket{\uparrow}_c\ket{\downarrow}_{c'}\bra{\downarrow}_c \bra{\uparrow}_{c'}
+\ket{\downarrow}_c\ket{\uparrow}_{c'}\bra{\uparrow}_c \bra{\downarrow}_{c'}\punc{.}
\end{aligned}
\end{equation}
Each term corresponds to one of the six ice-rules configurations of the tetrahedron, illustrated in \reffig{FigTetrGrid}, and is associated with a Boltzmann weight implementing the terms \(\scH_{\hv}\) and \(\scH_u\) in the Hamiltonian. (Terms breaking the ice rule could also be included; we are here treating such configurations as having zero Boltzmann weight.) The first four terms correspond to cases where the two chains maintain their configurations after passing through the tetrahedron---i.e., the two spins on each chain are aligned---while in the last two terms the chains exchange orientations.

Taking the inner product in \(\goH_{c'}\) gives
\begin{equation}
\scT\super{mf}_{c,z}
=\begin{aligned}[t]
&\left[\frac{1}{2}(1+m\super{mf}_{c'})\ee^{\frac{2}{\sqrt{3}}h/T}
+\frac{1}{2}(1-m\super{mf}_{c'})\ee^{2u/T}\right]\ket{\uparrow}_c\bra{\uparrow}_c
\\
&{}+ \left[ \frac{1}{2}(1-m\super{mf}_{c'})\ee^{-\frac{2}{\sqrt{3}}h/T}
+\frac{1}{2}(1+m\super{mf}_{c'})\ee^{2u/T}\right]\ket{\downarrow}_c\bra{\downarrow}_c\\
&{}+\zeta\super{mf}_{c'}\ket{\uparrow}_c\bra{\downarrow}_c
+\left({\zeta\super{mf}_{c'}}\right)^*\ket{\downarrow}_c\bra{\uparrow}_c\punc{,}
\end{aligned}
\label{EqTeff}
\end{equation}
where the expectation values
\begin{align}
m\super{mf}_{c'} &= \bra{v_{c'}}\big(\ket{\uparrow}\bra{\uparrow} - \ket{\downarrow}\bra{\downarrow}\big)\ket{v_{c'}}\\
\zeta\super{mf}_{c'} &= \bra{v_{c'}}\ket{\downarrow}\bra{\uparrow}\ket{v_{c'}}
\end{align}
are mean fields describing the magnetization and density of monopoles on chain \(c' = \scC_{c,z}\). The effective model for chain \(c\), described by the transfer operator \(\scT\super{mf}_{c,z}\), involves an applied field that is a function of \(m\super{mf}_{c'}\) and an effective monopole fugacity \(\zeta\super{mf}_{c'}\).

This is the crux of the approximation: The exchange of orientation between chains, described by the last two terms of \refeq{EqTetrahedronTransferOperator}, has been replaced by terms where a single chain flips orientation, thus creating or destroying a monopole. The criterion for deconfinement, that a distant pair of monopoles can be inserted with finite free-energy cost, is trivially satisfied whenever \(\zeta\super{mf}_{c'} \neq 0\). (In the analogous mean-field theory for the Bose--Hubbard model \cite{Fisher1989}, the expectation value of the bosonic annihilation operator is used as a mean field.)

In general, the solution of the self-consistent approximation requires finding simultaneous eigenvectors of \(\scT\super{mf}_{c,z}\), given in \refeq{EqTeff}, for all the values of \(z\). For the types of ordered phases expected in our model of spin ice, however, some further simplifications are possible. In particular, to describe the MDG and saturated phases, it is sufficient to divide the chains into two ``chain sublattices'', which we label \(1\) and \(2\), with one chain of each type meeting in every tetrahedron, as illustrated in \reffig{FigMDGChains}. The operator \(\scT\super{mf}_{c,z}\) is therefore independent of \(z\), and it is sufficient to find \(\ket{v_1}\) and \(\ket{v_2}\) such that \(\ket{v_c}_c\)  is the eigenvector of \(\scT\super{mf}_{c} = \bra{v_{c'}}_{c'} \scT_{cc'} \ket{v_{c'}}_{c'}\) with largest eigenvalue \cite{FootnotePerronFrobenius}, for both \(c = 1\), \(c'=2\) and vice versa. Because \(\scT\super{mf}_c\) is hermitian, this is equivalent to maximizing
\begin{equation}
\label{EqMatrixElement}
\Theta(\ket{v_1},\ket{v_2}) = \bra{v_{1}}_{1} \bra{v_{2}}_{2} \scT_{12} \ket{v_{1}}_{1}\ket{v_{2}}_{2}
\end{equation}
with respect to normalized \(\ket{v_1}\) and \(\ket{v_2}\).

As an aside, we note that this amounts to maximizing the matrix element of the layer transfer operator \(\scT_z\) in the subspace of factorizable vectors \(\ket{v} = \prod_c\ket{v_c}_c\). In these terms, it is clear how this is related to the mean-field theory for the Bose--Hubbard model, where one minimizes the matrix element of the Hamiltonian \(\scH\). (Recall the standard connection between classical and quantum statistical mechanics, where \(\ee^{-\tau\scH}\) is the evolution operator in imaginary time \(\tau\).) In both cases, the operator is hermitian and so the matrix element is bounded by its extremal eigenvalue, but only for the BH model is the Hamiltonian constant in time and the matrix element bounded by the true ground-state energy. In the present case, the product of extremal matrix elements \(\prod_z \bra{v}\scT_z \ket{v}\) provides an approximation to the exact partition function, but this approximation is not variational.

\begin{figure}
\begin{center}
\includegraphics{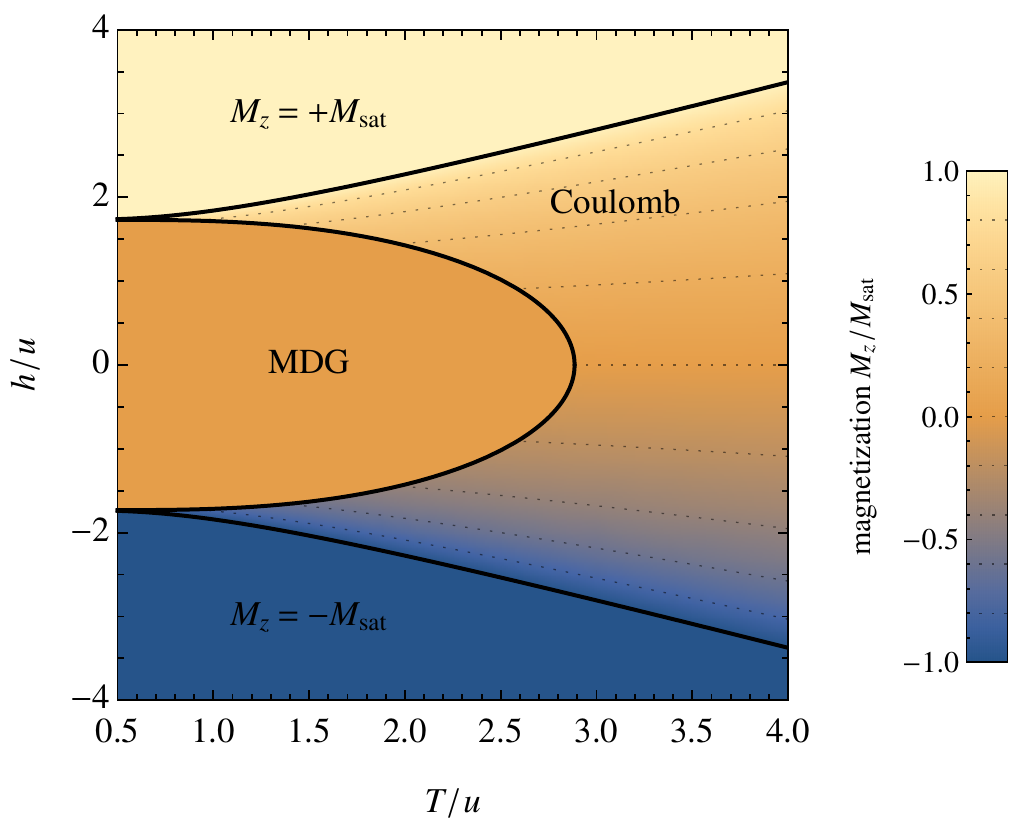}
\end{center}
\caption{Mean-field phase diagram, with magnetization shown on the color scale. The three low-temperature phases are, from top to bottom, a saturated paramagnet with \(M_z = +M\sub{sat}\), the Melko--den Hertog--Gingras (MDG) phase \cite{Melko2001} with \(M_z = 0\), and a saturated paramagnet with \(M_z = -M\sub{sat}\). At higher-temperatures is a Coulomb spin liquid whose magnetization is a smooth function of the applied field \(h\). (Dotted lines are contours of constant magnetization.)}
\label{FigPhaseDiagram}
\end{figure}

To maximize \(\Theta(\ket{v_1},\ket{v_2})\), we express both vectors as
\begin{equation}
\ket{v_c} = \ket{\uparrow} \cos \frac{\theta_c}{2} + \ket{\downarrow} \ee^{\ii \phi_c} \sin \frac{\theta_c}{2}\punc{,}
\end{equation}
where \(0 \le \theta_c \le \pi\) and \(-\pi \le \phi < \pi\), in terms of which
\begin{multline}
\label{EqTv1v2}
\Theta(\ket{v_1}, \ket{v_2}) = \frac{1}{2}(1+\cos\theta_1\cos\theta_2)\cosh \frac{2h}{\sqrt{3}T} + \frac{1}{2}(\cos\theta_1 + \cos\theta_2)\sinh \frac{2h}{\sqrt{3}T}\\
{}+\frac{1}{2}(1-\cos\theta_1\cos\theta_2)\ee^{2u/T}+\frac{1}{2}\sin\theta_1\sin\theta_2\cos(\phi_1 - \phi_2)\punc{.}
\end{multline}
The distinct maxima of this expression correspond to phases of the mean-field theory, shown in \reffig{FigPhaseDiagram}.
The phase boundaries can be found by expanding in small deviations from the ordered phases and identifying the points where they cease to be stable maxima. (We have verified numerically that the ordered states maximize the matrix element globally if and only if they are local maxima.)

At low temperature, there are three confined phases with different, fixed, values of the magnetization
\begin{equation}
\frac{M_z}{M\sub{sat}} = \frac{1}{2}\left(m\super{mf}_1 + m\super{mf}_2\right) = \frac{1}{2}(\cos\theta_1 + \cos\theta_2)
\punc{.}
\end{equation}
For \(\ee^{\frac{2}{\sqrt{3}}\lvert h\rvert/T} >\ee^{2u/T} + 1\), the maximum of \refeq{EqTv1v2} has \(\theta_c = 0\) (\(h > 0\)) or \(\theta_c = \pi\) (\(h < 0\)) for all \(c\), representing a fully polarized paramagnet with \(M_z = \pm M\sub{sat}\). For \(\ee^{u/T} > 2\cosh \frac{2h}{\sqrt{3}T}\), it has \(\theta_c = 0\) for \(c = 1\) and \(\theta_c = \pi\) for \(c = 2\), or vice versa. The two chain sublattices therefore have opposite orientations, \(m_1\super{mf} = -m_2\super{mf}\), and the total magnetization vanishes. Comparison with \reffig{FigMDG} confirms that this corresponds to the MDG phase.

At higher temperature, the maximum has \(0 < \theta_c < \pi\) for all \(c\) and \(\phi_1 = \phi_2\), corresponding to a paramagnet with continuously varying magnetization. This last phase is deconfined, having \(\zeta\super{mf}_c = \ee^{-\ii \phi_c}\sin\theta_c \neq 0\), and is therefore identified as the Coulomb phase.

\reffig{FigPhaseDiagram} also shows the magnetization \(M_z\) evaluated at the maximum of \(\Theta(\ket{v_1},\ket{v_2})\). The resulting phase diagram consists of a set of low-temperature confined phases with fixed magnetization, surrounded by the Coulomb phase in which the magnetization varies smoothly with the parameters. The MDG phase consists of a lobe, whose width, i.e., the range of fields for which it is stable, decreases with increasing temperature.

As a benchmark for the method, we note that our model reduces for \(u = 0\) to that considered by Jaubert et al.\ \cite{Jaubert2008} (in the limit of zero monopole fugacity). In this case, one can restrict to confined phases with all chains equivalent (i.e., \(\ket{v_1} = \ket{v_2}\)), and it is straightforward to show that the magnetization is given by
\begin{equation}
\frac{M_z}{M\sub{sat}} = \begin{cases}
\operatorname{sgn} h & \text{if \(T \le T\sub{K}\)}\\
\frac{\sinh \frac{2h}{\sqrt{3}T}}{2 - \cosh \frac{2h}{\sqrt{3}T}} & \text{if \(T > T\sub{K},\)}
\end{cases}
\end{equation}
where \(T\sub{K} = \frac{2h}{\sqrt{3}\log 2}\). Our method therefore agrees, in this case, with the results of the Bethe-lattice calculation \cite{JaubertThesis}.

\section{Discussion}
\label{SecPhaseStructure}

We have shown that applying mean-field theory to the model defined by \(\scH\) in \refeq{EqHamiltonian} produces a phase diagram containing the Coulomb phase \cite{Henley2010} as well as a fully-polarized paramagnet \cite{Jaubert2008} and the MDG phase \cite{Melko2001}. In this section, we argue that the general features in the phase diagram of \reffig{FigPhaseDiagram}, and in particular the lobe structure, can be understood by considering the interplay of confinement and flux in the Coulomb phase. We first consider the limit of vanishing monopole density, approximately valid at temperatures well below \(\Delta\), before discussing the qualitative effect of including monopoles.

\subsection{Structure of phase diagram}

The crucial observation underlying the phase structure is that confinement implies suppression of fluctuations in the magnetization \(\Mv\), and hence vanishing uniform magnetic susceptibility \(\chi\). By contrast, phases where monopoles are deconfined, such as the Coulomb phase, are magnetizable (i.e., have nonzero susceptibility \(\chi\)). This relationship between confinement and vanishing susceptibility (or, more generally, flux variance) occurs in any local model with a confining phase \cite{FootnoteMonopoleWinding}.

The lobe structure of the phase diagram follows immediately from this observation. Starting within a confining phase and changing the applied field will eventually tip the balance in favor of the Coulomb phase, whose free energy is quadratic in the field. The higher entropy in the Coulomb phase implies that the range of applied fields for which a given ordered state is favored narrows as the temperature increases. The confined phases therefore comprise lobes with fixed magnetization, each centered around favorable values of the applied field. (The fully polarized phases are exceptions, extending to arbitrarily large \(\lvert h \rvert\) as a result of their saturated magnetization.) This qualitative structure is confirmed in the mean-field phase diagram, \reffig{FigPhaseDiagram}. With more realistic interactions, we expect more lobes with other commensurate values of magnetization to appear in the gaps between the lobes shown, as indeed seen in the MC results of \refcite{Lin2013}.

The general phase structure can also be understood through the mapping to a quantum model of bosons, described in \refsec{SecBosonMapping} and \refcite{SpinIceCQ}. For an extended hard-core Bose--Hubbard model in the grand canonical ensemble, the phase diagram contains a set of Mott-insulator lobes at small hopping, consisting of different possible ordered states \cite{Fisher1989,Sachdev2011,Kovrizhin2005,Kurdestany2012}. Returning to spin ice, each of these corresponds to a confined phase with order in the spin degrees of freedom and fixed magnetization.

The susceptibility strictly vanishes in the ordered phases only in the thermodynamic limit at zero monopole fugacity \(\zeta = \ee^{-\Delta/T}\). With finite monopole cost, confinement of monopoles is no longer precisely defined and the susceptibility becomes nonzero, though remains exponentially small at low temperature. The plateaux are therefore rounded for nonzero \(\zeta\). Those phases that have conventional order, such as the MDG phase, remain separated from the paramagnet by a symmetry-breaking transition.

\subsection{Phase transitions}

Certain properties of the phase transitions can also be determined on general grounds, following similar arguments to those for the Bose--Hubbard model \cite{Fisher1989,Sachdev2011}, but with density replaced by magnetization. We first note that phase transitions can be classified by whether the magnetization changes across the transition. The magnetization necessarily changes in the case where the ordered phase has saturated magnetization, but it can also change across a transition into a magnetically ordered phase.

At the tip of the lobe, the magnetization is identical in the two phases (ordered within and Coulomb outside), and so this is the only point at which the magnetization stays the same across the transition. The argument is that, inside the ordered phase but sufficiently close to the tip, an arbitrarily small change in the applied field pushes the system into the Coulomb phase, regardless of its sign. If the two phases had different magnetizations at this point, their linear coefficients in the free energy would be different and this could not be the case. (This argument is essentially identical to the corresponding one for the Bose--Hubbard phase diagram \cite{Fisher1989}.)

For any continuous transitions with fixed magnetization, critical properties can be calculated using the approach of \refcite{SpinIceHiggs}, but with a modified projective symmetry group due to the nonzero magnetization. While MDG found a strongly first-order transition \cite{Melko2001} and the transition into the partially polarized phase was seen to be of first order in \refcite{Lin2013}, it remains possible that transitions in the presence of different perturbations could be continuous. (The possibility of unconventional transitions between a superfluid and fractional Mott insulators has been discussed in the extended Bose--Hubbard model \cite{Balents2005}.)

When the magnetization changes across the transition, i.e., everywhere except at the tip, the transition is in the same universality class as the Kasteleyn transition \cite{Kasteleyn1963,Jaubert2008}. It should be noted, however, that only the transitions into the saturated phases, with \(M_z = \pm M\sub{sat}\), are true Kasteleyn transitions, which are distinguished by the complete absence of fluctuations on the ordered side.

\section{Conclusions}
\label{SecConclusions}

We have presented a mean-field theory designed to study confinement transitions, based on the analogy between the confinement criterion and long-range order in conventional ordering transitions. The method has been applied to a simplified model of classical spin ice and shown to produce a phase diagram containing, besides the Coulomb spin liquid \cite{Henley2010}, a set of magnetization plateaux. These include the saturated paramagnet \cite{Jaubert2008} as well as the MDG phase expected to occur in dipolar spin ice at low temperature \cite{Melko2001}. We have argued that the general properties of the phase diagram follow from the interplay between confinement and magnetization that characterizes the ice-rule states and pointed out the analogy with the phase diagram of the extended Bose--Hubbard model \cite{Fisher1989}.

The simplified model that was studied here reproduces the MDG phase and is particularly amenable to the mean-field approach. It would be desirable to extend the approach to allow for more realistic further-neighbor interactions between spins, particularly the dipolar interactions that are known to be significant in the classical spin-ice materials. Further-neighbor interactions can also stabilize the partially magnetized phase observed using MC in \refcite{Lin2013}. (This phase can be described in terms of the chains shown in \reffig{FigMDGChains}, but distinguishes four chain sublattices, rather than two.)

A transfer operator written in the form of \refeq{EqTetrahedronTransferOperator} has the limitation, however, that it can represent interactions only within a single tetrahedron. (One cannot even define the layer transfer operator for general interactions.) A potential route to include further-neighbor interactions is through an additional mean-field decoupling, replacing a pairwise interaction \(V_{ij}\sigma_i\sigma_j\) with terms such as \(V_{ij}\sigma_i\langle\sigma_j\rangle\), representing a mean field acting on site \(i\). This extension, and others such as including nonzero monopole fugacity \(\zeta\), are left to future work.

\acknowledgments

This work was supported by EPSRC Grant No.\ EP/M019691/1.


\begin{thebibliography}{99}

\frenchspacing
\raggedright

\newcommand{\journal}[4]{#1 {\bf #2}, #3 (#4)}
\newcommand{\journaltitle}[5]{{\it #1}, \journal{#2}{#3}{#4}{#5}}
\newcommand{\journaltitlenovolume}[4]{{\it #1}, #2 ({\bf #4}), #3}
\newcommand{\book}[3]{{\it #1} (#2, #3)}
\newcommand{\prx}{Phys. Rev. X}
\newcommand{\arcmp}{Annu. Rev. Cond. Matt. Phys.}
\newcommand{\PR}[3]{\journal{\pr}{#1}{#2}{#3}}
\newcommand{\PRA}[3]{\journal{\pra}{#1}{#2}{#3}}
\newcommand{\PRB}[3]{\journal{\prb}{#1}{#2}{#3}}
\newcommand{\PRL}[3]{\journal{\prl}{#1}{#2}{#3}}
\newcommand{\PRX}[3]{\journal{\prx}{#1}{#2}{#3}}

\bibitem{Balents2010} L. Balents, \journaltitle{Spin liquids in frustrated magnets}{Nature}{464}{199}{2010}.

\bibitem{Henley2010} C. L. Henley, \journaltitle{The ``Coulomb phase'' in frustrated systems}{\arcmp}{1}{179}{2010}.

\bibitem{Bramwell2001} S. T. Bramwell and M. J. P. Gingras, \journaltitle{Spin ice state in frustrated magnetic pyrochlore materials}{Science}{294}{1495}{2001}.

\bibitem{Landau1999} L. D. Landau and E. M. Lifshitz, \book{Statistical Physics}{Butterworth--Heinemann, New York}{1999}.

\bibitem{Castelnovo2012} C. Castelnovo, R. Moessner, and S. L. Sondhi, \journaltitle{Spin ice, fractionalization, and topological order}{\arcmp}{3}{35}{2012}.

\bibitem{Castelnovo2008} C. Castelnovo, R. Moessner, and S. L. Sondhi, \journaltitle{Magnetic monopoles in spin ice}{Nature}{451}{42}{2008}.

\bibitem{MonopoleScalingPRL} S. Powell, \journaltitle{Universal monopole scaling near transitions from the Coulomb phase}{\prl}{109}{065701}{2012}.

\bibitem{MonopoleScalingPRB} S. Powell, \journaltitle{Confinement of monopoles and scaling theory near unconventional critical points}{\prb}{87}{064414}{2013}.

\bibitem{Jaubert2008} L. D. C. Jaubert, J. T. Chalker, P. C. W. Holdsworth, and R. Moessner, \journaltitle{Three-dimensional Kasteleyn transition: Spin ice in a \([100]\) field}{\prl}{100}{067207}{2008}.

\bibitem{SpinIceCQ} S. Powell and J. T. Chalker, \journaltitle{Classical to quantum mappings for geometrically frustrated systems: Spin-ice in a \([100]\) field}{\prb}{78}{024422}{2008}.

\bibitem{SpinIceHiggs} S. Powell, \journaltitle{Higgs transitions of spin ice}{\prb}{84}{094437}{2011}.

\bibitem{Fisher1989} M. P. A. Fisher, P. B. Weichman, G. Grinstein, and D. S. Fisher, \journaltitle{Boson localization and the superfluid-insulator transition}{\prb}{40}{546}{1989}.

\bibitem{Sachdev2011} S. Sachdev, \book{Quantum Phase Transitions}{Cambridge University Press, Cambridge, England}{2011}.

\bibitem{Matsuhira2002} K. Matsuhira, Z. Hiroi, T. Tayama, S. Takagi, and T. Sakakibara, \journaltitle{A new macroscopically degenerate ground state in the spin ice compound \DTO\ under a magnetic field}{J. Phys.: Condens. Matter}{14}{L559}{2002}.

\bibitem{Moessner2003} R. Moessner and S. L. Sondhi, \journaltitle{Theory of the \([111]\) magnetization plateau in spin ice}{\prb}{68}{064411}{2003}.

\bibitem{Isakov2004} S. V. Isakov, K. S. Raman, R. Moessner, and S. L. Sondhi, \journaltitle{Magnetization curve of spin ice in a \([111]\) magnetic field}{\prb}{70}{104418}{2004}.

\bibitem{Shannon2012} N. Shannon, O. Sikora, F. Pollmann, K. Penc, and P. Fulde, \journaltitle{Quantum ice: a quantum Monte Carlo study}{\prl}{108}{067204}{2012}.

\bibitem{Sreejith2014} G. J. Sreejith and S. Powell, \journaltitle{Critical behavior in the cubic dimer model at nonzero monomer density}{\prb}{89}{014404}{2014}.

\bibitem{FootnoteTotalChargeZero} Other terms vanish because they correspond to configurations where the total charge is nonzero. We assume (spin-inversion or spatial) symmetry so that \(G\sub{m}(\rv) = G\sub{m}(-\rv)\).

\bibitem{Yeomans1992} J. M. Yeomans, \book{Statistical Mechanics of Phase Transitions}{Oxford University Press, Oxford}{1992}.

\bibitem{Henelius2016} P. Henelius, T. Lin, M. Enjalran, Z. Hao, J. G. Rau, J. Altosaar, F. Flicker, T. Yavors'kii, and M. J. P. Gingras, \journaltitle{Refrustration and competing orders in the prototypical \DTO\ spin ice material}{\prb}{93}{024402}{2016}.

\bibitem{Melko2001} R. G. Melko, B. C. den Hertog, and M. J. P. Gingras, \journaltitle{Long-range order at low temperatures in dipolar spin ice}{\prl}{87}{067203}{2001}.

\bibitem{FootnoteNomenclature} This nomenclature follows Lin and Kao \cite{Lin2013}. McClarty et al.\ \cite{McClarty2015} call the same state a ``cubic antiferromagnet'', while Henelius et al.\ \cite{Henelius2016} call it the ``single-chain state''. In \refcite{SpinIceHiggs}, it was called a ``spin spiral''. MDG themselves \cite{Melko2001} refer to it by its ordering wavevector, \(\qv =\frac{2\pi}{a}\uvz\), where \(a = 2\sqrt{2}\rnn\) is the side length of the cubic unit cell.

\bibitem{Lin2013} S.-C. Lin and Y.-J. Kao, \journaltitle{Half-magnetization plateau of a dipolar spin ice in a \([100]\) field}\PRB{88}{220402(R)}{2013}.

\bibitem{McClarty2015} P. A. McClarty, O. Sikora, R. Moessner, K. Penc, F. Pollmann, and N. Shannon, \journaltitle{Chain-based order and quantum spin liquids in dipolar spin ice}{\prb}{92}{094418}{2015}.

\bibitem{BrooksBartlett2014} M. E. Brooks-Bartlett, S. T. Banks, L. D. C. Jaubert, A. Harman-Clarke, and P. C. W. Holdsworth, \journaltitle{Magnetic-moment fragmentation and monopole crystallization}{\prx}{4}{011007}{2014}.

\bibitem{Reimers1991} J. N. Reimers, A. J. Berlinsky, and A.-C. Shi, \journaltitle{Mean-field approach to magnetic ordering in highly frustrated pyrochlores}{\prb}{43}{865}{1991}.

\bibitem{Baxter} R. J. Baxter, \journaltitle{Dimers on a rectangular lattice}{J. Math. Phys.}{9}{650}{1968}.

\bibitem{Rujan} P. Ruj\'an, \journaltitle{Variational method for lattice systems: General formalism and application to the two-dimensional Ising model in an external field}{Physica}{96A}{379}{1979}.

\bibitem{Bukov2015} M. Bukov, L. D'Alessio, and A. Polkovnikov, \journaltitle{Universal high-frequency behavior of periodically driven systems: from dynamical stabilization to Floquet engineering}{Adv. Phys.}{64}{139}{2015}.

\bibitem{FootnotePerronFrobenius} The effective transfer matrix \(\scT\super{mf}_c\) has all positive elements, and so, by the Perron--Frobenius theorem, the largest eigenvalue by absolute value is positive.

\bibitem{JaubertThesis} L. D. C. Jaubert, Ph.D.\ thesis, \'Ecole Normale Sup\'erieure de Lyon, 2009 [\url{http://hal.archives-ouvertes.fr/docs/00/46/29/70/PDF/Thesis.pdf}].

\bibitem{FootnoteMonopoleWinding} One can imagine creating a monopole--antimonopole pair, winding one of the pair around the periodic boundaries, and then recombining them \cite{Hermele2004}. The result of this charge transport is to increase the flux through the system; since the separation required diverges in the thermodynamic limit, it is possible with finite energy cost only in a deconfined phase.

\bibitem{Hermele2004} M. Hermele, M. P. A. Fisher, and L. Balents, \journaltitle{Pyrochlore photons: The \(\rmU(1)\) spin liquid in a \(S=1/2\) three-dimensional frustrated magnet}{\prb}{69}{064404}{2004}.

\bibitem{Kovrizhin2005} D. L. Kovrizhin, G. Venketeswara Pai, and S. Sinha, \journaltitle{Density wave and supersolid phases of correlated bosons in an optical lattice}{EPL}{72}{162}{2005}.

\bibitem{Kurdestany2012} J. M. Kurdestany, R. V. Pai, and R. Pandit, \journaltitle{The inhomogeneous extended Bose-Hubbard model: A mean-field theory}{Annalen der Physik}{524}{234}{2012}.

\bibitem{Balents2005} L. Balents, L. Bartosch, A. Burkov, S. Sachdev, and K. Sengupta, \journaltitle{Putting competing orders in their place near the Mott transition}{\prb}{71}{144508}{2005}.

\bibitem{Kasteleyn1963} P. W. Kasteleyn, \journaltitle{Dimer statistics and phase transitions}{J. Math. Phys.}{4}{287}{1963}.

\end{thebibliography}
\end{document}